# Optimizing Metal-Organic Chemical Vapor Deposition for Ultrawide-Band-Gap MgSiN$_2$ Thin Films

Chenxi Hu[1,*], Abdul Mukit[2,*], Vijay Gopal Thirupakuzi Vangipuram[2], Christopher Chae[3], Jinwoo Hwang[3], Kathleen Kash[1], Hongping Zhao[2,3,†]

[1]Department of Physics, Case Western Reserve University, Cleveland OH 44106
[2]Department of Electrical and Computer Engineering, The Ohio State University, Columbus OH 43210
[3]Department of Materials Science and Engineering, The Ohio State University, Columbus OH 43210
[*] Equally contributed first authorship
[†] Corresponding author: zhao.2592@osu.edu

## Abstract

Orthorhombic II–IV nitride semiconductors offer an expanded and more tunable material set with unique properties, while maintaining close compatibility with the wurtzite crystal structure of the III-nitrides. In particular, MgSiN$_2$—a II–IV nitride closely lattice matched to GaN and AlN—has a band gap suitable for photonic applications in the UV-C wavelength region. MgSiN$_2$ is also a promising candidate to exhibit ferroelectricity, which has only been observed in very few nitride materials. This study builds on our previous work on the metal-organic chemical vapor deposition (MOCVD) of MgSiN$_2$ thin films grown on GaN-on-sapphire and c-plane sapphire substrates by exploring higher growth temperature windows, resulting in higher crystalline quality and improved interfaces. Correlations between the growth conditions (Mg:Si precursor molar flow rate ratio, reactor pressure, and growth temperatures from 900 °C to 960 °C) and the resultant film quality are investigated for films grown on GaN-on-sapphire. High-resolution transmission electron microscopy (HR-TEM) reveals high-quality orthorhombic single-crystal MgSiN$_2$, confirming successful epitaxial growth on GaN. Optical transmittance measurements indicate the direct band gap is 6.34-6.36 eV and indirect band gap is 5.77-5.81 eV, affirming the realization of an ultrawide-band gap II–IV nitride semiconductor that is structurally compatible with existing III-nitride device platforms.

# I. Introduction

In recent years, heterovalent II–IV–N$_2$ compounds have emerged as a compelling extension of the III–N semiconductors, thanks to their structural and chemical affinity with wurtzite GaN, AlN, and InN. Conceptually, the II-IV-nitride materials can be derived from the III–N lattice by substituting each pair of group-III atoms with one group-II atom and one group-IV atom, such that every nitrogen atom remains coordinated to two cations of each type [1-3]. This structural motif preserves many advantageous III–N features—such as robust chemical stability—yet also introduces new fundamental properties. Band gap tuning via cation-sublattice disorder becomes possible in II–IV–N$_2$ systems without modifying their stoichiometry. The formation of heterostructures between the II-IV-N$_2$ and the III-N materials with close lattice-matching but large band offsets become feasible.

Previously, related II–IV–N$_2$ materials such as ZnGeN$_2$ and ZnSnN$_2$ were grown in thin-film form by metal-organic chemical vapor deposition (MOCVD) [4-5]. Additionally, MOCVD studies have shown the successful growth of MgGeN$_2$ thin films, grown at a pyrometer temperature of 745 °C, that yielded a band gap of $4.28 \pm 0.06$ eV [6]. Among the various II–IV–N$_2$ compounds, MgSiN$_2$ stands out for its ultrawide band gap (theoretically predicted at 6.28 eV direct and 5.84 eV indirect) [7-9]. This large band gap suggests a strong potential for next-generation deep-UV optoelectronics and high-power electronics, especially when paired with the close lattice match of MgSiN$_2$ to GaN and AlN. Like other II–IV–N$_2$ compounds, the presence of two cations in MgSiN$_2$ adds a new axis of tunability for material properties like the band gap and band offsets. These properties can be tuned by alloying with other II-IV-nitride compounds and with the III-nitrides.

Tuning of properties by alloying within the pure III-N compounds is limited in its range by the lattice mismatch among these compounds [10].

Previous theoretical studies of MgSiN$_2$ predicted that it should exhibit excellent mechanical stability under high pressure (up to 10 GPa) and large in-plane elastic coefficients, dielectric, and specific heat capacity at constant volume [9,11]. Moreover, MgSiN$_2$ is predicted to have attractive ferroelectric properties, with a stronger breakdown field and a lower coercive field than those predicted for AlN [12,13].

MgSiN$_2$ thin films have been realized using a reactive sputtering process. These films show a non-equilibrium wurtzite structure due to cation disordering [14]. Our previous study on the growth of MgSiN$_2$, in which we explored relatively low to moderate temperature growth windows (745°C~850°C), produced films that displayed good crystallinity [15]. Although the experiments exhibited promising results, it also suggested that improvements in crystallinity and interface quality could be made.

Deposition of high quality ultra-wide bandgap materials generally requires higher growth temperatures for optimal adatom diffusion, cation ordering and nitrogen incorporation. In this work, we build upon our previous work and extend the MOCVD growth window for MgSiN$_2$ thin films into a higher temperature regime ($\geq$ 900 °C). We investigate how precursor molar flow rate ratios, ammonia flow rate, growth temperature, and reactor pressure influence film quality, crystallinity, surface morphology, and optical properties. The MgSiN$_2$ films grown under the optimal conditions achieve higher crystal quality, as evaluated by x-ray diffraction, transmission electron microscopy, atomic force microscopy, and optical transmittance, different characterization techniques and an ultrawide band gap in line with theoretical predictions. Our results thus highlight the feasibility of integrating MgSiN$_2$ into existing III–N processing platforms and illustrate a clear path toward ultrawide-band gap II–IV–N$_2$ devices for deep-UV and high power-density applications.

## II. Experimental Details

MgSiN$_2$ thin films were grown in a custom vertical rotating-disk MOCVD reactor, using high-purity N$_2$ as the carrier gas. Silane (SiH$_4$) served as the Si precursor, while bis(methylcyclopentadienyl)magnesium (MeCp)$_2$Mg was chosen as the Mg precursor due to its relatively higher vapor pressure compared to the more commonly used Cp$_2$Mg. The bubbler temperature for (MeCp)$_2$Mg was maintained at 40°C. At this temperature, the molar flow rate for the Mg precursor was estimated to be 9.7 µmol/min. The SiH$_4$ flow rate was adjusted to vary the Mg:Si molar ratio as needed to obtain good stoichiometry.

All MgSiN$_2$ films in this study were grown on GaN-on-sapphire (GaN/c-sapphire) templates, with threading dislocations densities of ~ 4×10$^8$ cm$^{-2}$, co-loaded with double-side-polished c-sapphire substrates.

As summarized in Table 1, multiple series of MgSiN$_2$ samples were grown under systematically varied conditions to evaluate the influences of growth temperature, reactor pressure, and Mg:Si molar flow rate ratio (FRR). For samples A–C, the growth temperature and pressure were fixed at 900 °C and 500 Torr, respectively, while the Mg:Si FRR and NH$_3$ flow were adjusted. Samples D–F were also grown at 500 Torr but with the NH$_3$ flow fixed at 2.5 slm and the temperature varied from 928 °C to 960 °C. Further exploration at 945 °C (with 2.5 slm NH$_3$ flow rate) was divided into two subsets: Samples E and G–J, grown at 500 Torr, and Samples K–O, grown at 450 Torr. Samples within each series were grown with different Mg:Si ratios to isolate the effects of pressure and precursor composition. All growth runs lasted 1 hour, yielding film thicknesses typically in the 50–150 nm range. Throughout the experiments, the $N_2$ carrier gas flow was maintained at 2000 sccm.

A Thermo Scientific Apreo scanning electron microscopy (SEM) was used at an acceleration voltage of 30 keV for imaging and 5 keV for energy dispersive X-ray spectroscopy to assess surface morphology and elemental composition. Atomic force microscopy (AFM) was done with a Bruker AXS Dimension Icon system in tapping mode over 5 µm × 5 µm scan areas. Measurements were performed on representative regions across the wafer to evaluate surface uniformity. A Bruker D8 Discover using Cu Kα radiation ($\lambda$ = 1.5416 Å) was employed for X-ray diffraction (XRD) ω–2θ scans and rocking curves. High-resolution transmission electron microscopy (HR-TEM) imaging was performed on a Thermo Fisher probe-corrected Titan STEM. An Agilent Cary 5000 UV–Vis–NIR spectrophotometer was used to measure optical transmittance from 1.0 eV to 6.7 eV. Band gap energies were extracted by Tauc plot analyses and compared to theoretical predictions.

### III. Results and Discussion

Section A studies how variations in the II–IV molar FRR and $NH_3$ flow rate affect the morphological and structural properties of the $MgSiN_2$ films. In section B we examine the effects of growth temperature on the growth process. Section C assesses the influence of different reactor pressures on the optimization of FRRs to obtain stoichiometry. In section D, we discuss the results of transmittance measurements to extract the direct band gap. Section E discusses a detailed TEM analysis to evaluate the crystalline quality and interface characteristics of these films.

### III.A. Effects of variations in II-IV molar FRRs and $NH_3$ flow rates

Figure 1 gives an overview of the effect of the MOCVD growth conditions on the stoichiometry of the films. The trend shows that to maintain near-stoichiometric films at higher growth temperatures, a higher Mg:Si FRR is required. In addition, we see that at a given temperature reducing the reactor pressure decreases the Mg:Si atomic ratio, indicating that the incorporation of magnesium diminishes under lower-pressure conditions. These phenomena are discussed in more detail in sections A, B, and C.

To evaluate the interplay between Mg:Si precursor molar flow rate ratios and $NH_3$ flow at higher growth temperatures, a series of $MgSiN_2$ films was grown under different combinations of these parameters. Figure 2 illustrates how varying the Mg:Si molar FRR and $NH_3$ flow rate influence the film morphology, crystallinity, and stoichiometry for samples A-C, grown at 900 °C. We see that the dominant diffraction peaks in the XRD spectra near 34.5° originate from the underlying (002) GaN layer, while the $MgSiN_2$ peak typically appears around 36–37°, shifting in position and intensity according to the Mg:Si ratio and $NH_3$ flow rate. Notably, the position of the strongest $MgSiN_2$ peak —close to 36.1°, seen in sample A —corresponds well with the theoretical (002) $MgSiN_2$ reflection, suggesting near-ideal stoichiometry and minimal residual strain.

Sample A was grown with an $NH_3$ flow of 2.5 slm and a Mg:Si FRR of 9. Although the film exhibits a large root-mean-square (RMS) roughness, its XRD spectrum reveals a pronounced and relatively narrow $MgSiN_2$ peak, indicative of good crystallinity. The better crystalline quality could be attributed to the sufficient availability of the nitrogen precursor, which is further promoted by the higher cracking rate of ammonia at higher temperatures. This sufficient $NH_3$ concentration would promote the lowering of nitrogen vacancy sites resulting in fewer dislocations and stronger XRD peaks. For sample B the $NH_3$ flow was increased to 3.5 slm while the Mg:Si FRR was held constant. The increase in the $NH_3$ flow rate resulted in coarser grain structures, a broader, slightly shifted $MgSiN_2$ peak, and a more Mg-rich film. For sample C the $NH_3$ flow rate was maintained at 3.5 slm while the $SiH_4$ flow rate was increased in order to decrease the Mg:Si FRR from 9 to 7.5. Although the resulting RMS roughness and composition are comparable to those of sample A, the XRD peak is broader and shifted away from the ideal $MgSiN_2$ position, possibly indicating the incorporation of strain into the film. As observed, higher ammonia flow reduces Si incorporation, which could result in some Mg-rich domains and defects, also contributing to the shift of the XRD peak.

From these observations, an $NH_3$ flow rate of 2.5 slm appears most advantageous for high-temperature $MgSiN_2$ growth, as it consistently produces relatively smooth films with a strong (002) $MgSiN_2$ diffraction peak. Accordingly, this $NH_3$ flow condition was used for subsequent sample series in this study.

### III.B. Effects of variations in temperature

To explore how growth temperature influences the $MgSiN_2$ film quality, a series of samples was grown at 900 °C (sample A), 928 °C (sample D), 945 °C (sample E), and 960 °C (sample F). Mg:Si FRRs were increased with temperature in order to maintain good stoichiometry. The AFM, SEM and XRD scans for these samples are shown in Figure 3. At 900 °C, sample A's moderate Mg:Si (FRR) of 9.0 yielded a slightly Mg-rich film composition ($1.12 \pm 0.10$ from SEM-EDX) and relatively high RMS roughness (~24.2 nm). Increasing the temperature to 928 °C (sample D) improved film uniformity, lowering the roughness to ~16.2 nm and moving the composition closer to a stoichiometric Mg:Si ratio of $1.03 \pm 0.10$. As shown in Figure 3, the $MgSiN_2$ diffraction peak for sample D is slightly stronger and narrower than in sample A, indicating an improvement in crystal quality. Pushing the temperature to 945 °C (sample E) the lowest RMS roughness (~5 nm) was achieved for this series. These findings suggest that the optimal growth window for high-quality $MgSiN_2$ films lies between 945 °C and 960 °C, where a favorable balance between precursor decomposition, surface diffusion, and stoichiometric control can be achieved. Higher reactor temperature generally translates to enhanced surface migration of adatoms, efficient cracking of ammonia and increased surface reaction rate, resulting in smoother morphology and better crystallinity [16, 17, 18]. However, desorption of Mg increases with temperature [19], which necessitates higher Mg:Si molar FRR to achieve stoichiometry. Although sample F (grown at 960 °C) exhibited a broader $MgSiN_2$ peak—indicative of possible off-stoichiometry composition or residual strain—this broadening may be partly due to slight variations in the Mg:Si flow rate ratio rather than an inherent limitation of the higher temperature. With further fine-tuning of the FRR, films grown at 960 °C might achieve comparable crystallinity and stoichiometry to those grown at 945 °C. In short, our results indicate that the growth window for optimal material quality can be defined within the 945–960 °C range. As the sample E grown at 945 °C exhibited comparatively better surface morphology than sample F, the growth temperature of 945 °C was selected for the next batch of experiments.

### III.C. Effects of changes in the FRR and the reactor pressure

We next investigate how the reactor pressure and the Mg:Si flow rate ratio (FRR) together influence film stoichiometry and crystallinity. Three series of samples were grown at 945 °C with different FRRs; at 500 Torr (samples G-J), at 450 Torr (samples K-O), and at 550 Torr (samples P-R). These three series are included in the graph of composition versus FRR in Figure 1. In examining the series grown at 500 Torr, as shown in Figure 4, we see that sample G was grown with an especially high $SiH_4$ flow—resulting in a low FRR—which led to a highly off-stoichiometry Mg-poor film with a Mg:Si ratio of $0.57 \pm 0.05$. XRD reveals two secondary peaks to the right of the main $MgSiN_2$ reflection. Reducing the $SiH_4$ flow by about 15% (sample E) resulted in good stoichiometry (a Mg:Si ratio of $0.99 \pm 0.09$) and a single narrow $MgSiN_2$ peak. Interestingly, further reducing the $SiH_4$ flow rate sequentially in samples H, I, and J results in films with just as good stoichiometry, within the uncertainties of the EDX measurements. We can infer that there is a driving force toward ideal stoichiometry at higher FRRs, perhaps arising from a tendency for excess Mg adsorbed on the surface to evaporate. Sample I shows the strongest, sharpest MgSiN$_2$ diffraction peak alongside a remarkably low RMS roughness of $1.5 \pm 0.1$ nm. However, pushing the FRR even higher for sample J resulted in increased roughness.

Similar trends emerged at 450 Torr, as shown in Figure 5 for samples K through O. Sample K was again Mg-poor and exhibited a distinct secondary XRD peak. Raising the molar FRR to get good stoichiometry caused the secondary peak to disappear and resulted in smoother, well-crystallized films—best exemplified by sample N, which had an RMS roughness of $3.1 \pm 0.2$ nm. Excessively high FRRs, on the other hand, pushed the films toward Mg-rich conditions, leading to surface roughening and 3D growth features (sample O). Under lower pressure the adatoms are uniformly dispersed, which results in a smoother and more uniform surface [20]. However, reaching stoichiometry at 450 Torr requires a higher Mg flow than at 500 Torr. This result is consistent with greater Mg evaporation from the surface at lower chamber pressures and/or a lower efficiency for Mg incorporation at the lower pressure.

In the 550 Torr series (Figure 6), a similar trend is again observed; for instance, sample P exhibits a shoulder on the right side of the MgSiN$_2$ peak—indicative of an off-stoichiometric phase—even though its peak intensity is similar to that of stoichiometric films grown at 450 Torr and 500 Torr, and the SEM-EDS, taking into account the uncertainty, also suggests that the film should be near-stoichiometric. As the Mg:Si molar FRR is increased, it is observed that surface morphology improves, and XRD peaks become narrower, indicating better crystal quality. However, the sample showing the sharpest peak (sample R) in the 550 Torr series is rougher compared to the results for 450 and 500 torr. The increase in chamber pressure raises the concentration of reactant species, causing faster growth rates while also extending the residence period and increasing parasitic reaction rates [21]. These effects may help explain the higher roughness observed at 550 torr. These observations indicate that both reactor pressure and FRR can be optimized to achieve near-stoichiometric $MgSiN_2$ films with minimal roughness and high crystallinity. A chamber pressure of 500 Torr requires a higher Si flow to avoid Mg-poor phases than a chamber pressure of 450 Torr. At 550 Torr, even close-to-stoichiometric films tend to exhibit rougher surfaces than at lower pressures. These findings reveal the intricate interplay between gas-phase reactions, surface kinetics, and precursor delivery in determining the film properties.

Following these growth optimizations, high-resolution rocking curve measurements of the (002) $MgSiN_2$ reflections provide additional insight into crystalline quality. In Figure 7, sample I, which exhibits minimal surface roughness and a well-defined $MgSiN_2$ peak in the $\omega$–$2\theta$ scan, shows a relatively narrow FWHM (198 ± 2 arcsec) in its rocking curve. This narrow FWHM indicates reduced mosaicity and improved crystallographic alignment, consistent with the near-stoichiometric composition and smooth morphology observed via AFM and SEM. These data demonstrate that fine-tuning both the FRR and reactor pressure at a growth temperature of ~ 945 °C can yield high-quality $MgSiN_2$ films with low defect densities.

### III.D. Optical band gap extraction

To measure their transmittance, films were grown on double-side-polished c-plane sapphire substrates under various stoichiometric conditions (detailed in Table 2). The samples include a Mg-rich film (sample A), a Mg-poor film (sample K), and near-stoichiometric films (samples D, E, and F). The stoichiometric films exhibiting notably lower RMS roughness than the Mg-rich and Mg-poor films.

Figure 8(a) plots the transmittance spectra for these samples. Tauc plot analyses, shown in Figures 8(b) and 8(c), were performed to extract both the direct and indirect band gaps by extrapolating the linear regions of $(\alpha h\nu)^2$ versus $h\nu$ and $(\alpha h\nu)^{1/2}$ versus $h\nu$, respectively. For the direct band gap determination, the linear fits were carried out over a photon energy range of 6.5–6.65 eV, while for the indirect band gap, the fitting range was 6.3–6.5 eV. These ranges were chosen by considering the predicted band structure. [8] For the near-stoichiometric films, the extracted direct band gaps were $6.35 \pm 0.03$ eV (sample D), $6.38 \pm 0.03$ eV (sample F), and $6.36 \pm 0.03$ eV (sample E), which agree well with the theoretical value of 6.28 eV. The indirect band gaps for these samples ranged from $5.77 \pm 0.03$ eV to $5.81 \pm 0.03$ eV, closely matching the predicted 5.84 eV. Transmittance data for sample A (Mg-rich) and sample K (Mg-poor) reveal lower direct and indirect band gaps compared to the near-stoichiometric films, indicating that band gap measurements can be correlated with crystal stoichiometry and quality. These measurements are summarized in Table 2.

### III.E. Scanning Transmission Electron Microscopy

High-angle annular dark field (HAADF) STEM imaging was used to investigate sample E. Figure 9 (a) and (b) show two magnifications of the cross-section of the film, where the clear atomic structure of $MgSiN_2$ is seen near the GaN interface and up to 40 nm away from the interface, indicating uniform epitaxial growth of the film. The $MgSiN_2$/GaN interface is uniform across the imaged areas, showing no clear signs of defects originating from the interface. STEM EDX was used to obtain an elemental map of the cross-sectional area of the $MgSiN_2$ film and across the film-substrate interface. Figure 9 (c) shows the area over which the signal was collected, with the highlighted box indicating the area over which the averaged line scan was taken. Figure 9 (d) shows the individual elemental maps. The quantitative EDX measurements are plotted in Fig. 9(e).

The thickness of sample E is approximately 70 nm. The averaged line scan shows a consistent composition throughout the film. Table 3 summarizes the extracted atomic fractions in the expected 1:1:2 ratio for stoichiometric $MgSiN_2$, within measurement uncertainty.

Electron diffraction patterns were also taken from the $MgSiN_2$ film region and the GaN template region to examine the crystal structure and uniformity of the $MgSiN_2$ film growth. Figures 10 (a) and (b) show the simulated diffraction pattern for GaN and the experimental diffraction pattern collected from STEM, which closely matches the simulated pattern, as expected. The simulated diffraction pattern for orthorhombic $MgSiN_2$ is compared to the experimental diffraction pattern collected from STEM in Figures 10 (c) and (d), respectively. To minimize the signal from the GaN template beneath the $MgSiN_2$ film, the diffraction pattern was collected in the middle of the film region. The measured diffraction pattern is clearly orthorhombic, as one expects for substantial ordering on the cation sublattice, instead of wurtzitic, as would be expected for substantial amounts of disorder on the cation sublattice.

## IV. Conclusions

We have demonstrated that $MgSiN_2$ film quality and stoichiometry are highly sensitive to the II–IV molar flow rate ratios, the $NH_3$ flow rate, the growth temperature, and the reactor pressure. Inadequate $NH_3$ flows or imbalanced Mg:Si ratios produced Mg-rich or Mg-poor films, as evidenced by the $MgSiN_2$ (002) XRD peak intensities and the higher RMS roughness values measured by AFM. In contrast, moderately higher $NH_3$ flow (2.5 slm) and balanced Mg:Si ratios yielded near-stoichiometric $MgSiN_2$ with a particularly stronger and narrower (002) diffraction peak near 36.7°, minimal surface roughness, and low defect densities as confirmed by XRD rocking curve FWHM measurements. Growth temperature studies revealed 945 °C as the optimal growth temperature for MgSiN$_2$ growth, providing an ideal balance of precursor decomposition, adatom mobility, and stoichiometric control. The STEM diffraction pattern of a representative MgSiN$_2$ films showed an orthorhombic crystal structure. Finally, optical transmittance measurements of films grown on transparent sapphire confirmed an ultrawide direct band gap of 6.34-6.36 eV and an indirect band gap of 5.77-5.81 eV, consistent with the theoretical predictions. Together, these results establish a clear set of process parameters—centered around 945 °C and carefully tuned pressures and Mg:Si molar FRR—for achieving smooth, stoichiometric MgSiN$_2$ layers suitable for device applications. Compared to our previous work, these results demonstrate that high-temperature MOCVD growth (in the 945–960 °C range) significantly broadens the growth window for achieving high-quality, near-stoichiometric MgSiN$_2$ films. The refined control over Mg:Si flow ratios and reactor pressure has enabled improved film crystallinity and interface quality, marking a notable advancement toward robust integration of MgSiN$_2$ with conventional III-nitride platforms.

**Acknowledgements**

This work was supported by the Army Research Office (Award No. W911NF-24-2-0210).

**Conflict of Interest**

The authors have no conflicts of interest to disclose.

**Data Availability**

The data that support the findings of this study are available from the corresponding author upon reasonable request.

# References

[1]. Lambrecht, W. R.; Punya, A. Heterovalent Ternary II-IV-N$_2$ Compounds: Perspectives for a New Class of Wide-Band-Gap Nitrides. III-Nitride Semiconductors and their Modern Devices 2013, 519–585.

[2]. Lyu, S.; Lambrecht, W. R. Band Alignment of III-N, ZnO and II–IV-N$_2$ Semiconductors from the Electron Affinity Rule. Journal of Physics D: Applied Physics 2019, 53 (1), 015111.

[3]. Lyu, S.; Skachkov, D.; Kash, K.; Blanton, E. W.; Lambrecht, W. R. Band Gaps, Band-Offsets, Disorder, Stability Region, and Point Defects in II-IV-N2 Semiconductors. Physica Status Solidi (a) 2019, 216 (15), 1800875.

[4]. Karim, M. R.; Jayatunga, B. H. D.; Feng, Z.; Kash, K.; Zhao, H. Metal–Organic Chemical Vapor Deposition Growth of ZnGeN2 Films on Sapphire. Crystal Growth & Design 2019, 19 (8), 4661–4666.

[5]. Karim, M. R.; Jayatunga, B. H. D.; Zhu, M.; Lalk, R. A.; Licata, O.; Mazumder, B.; Hwang, J.; Kash, K.; Zhao, H. Effects of Cation Stoichiometry on Surface Morphology and Crystallinity of ZnGeN2 Films Grown on GaN by Metalorganic Chemical Vapor Deposition. AIP Advances 2020, 10 (6).

[6]. Hu, C.; Vangipuram, V. G. T.; Chae, C.; Turan, I. K.; Hoven, N.; Lambrecht, W. R.; Zhao, H; Kash, K. Metal-organic chemical vapor deposition of MgGeN$_2$ films on GaN and sapphire. arXiv preprint arXiv:2502.18618.

[7]. Lyu, S.; Lambrecht, W. R. Quasiparticle Self-Consistent GW Band Structures of Mg-IV-N2 Compounds: The Role of Semicore d States. Solid State Communications 2019, 299, 113664.

[8]. Jaroenjittichai, A. P.; Lambrecht, W. R. Electronic Band Structure of Mg-IV-N2 Compounds in the Quasiparticle-Self-Consistent GW Approximation. Physical Review B 2016, 94 (12), 125201.

[9]. Råsander, M.; Quirk, J.; Wang, T.; Mathew, S.; Davies, R.; Palgrave, R.; Moram, M. Structure and Lattice Dynamics of the Wide Band Gap Semiconductors MgSiN2 and MgGeN2. Journal of Applied Physics 2017, 122 (8).

[10]. Holec, D.; Zhang, Y.; Rao, D. V. S.; Kappers, M. J.; McAleese, C.; Humphreys, C. J. Equilibrium Critical Thickness for Misfit Dislocations in III-Nitrides. Journal of Applied Physics 2008, 104 (12), 123514.

[11]. Bootchanont, A., Phacheerak, K., Fongkaew, I., Limpijumnong, S., & Sailuam, W. (2021). The pressure effect on the structural, elastic, and mechanical properties of orthorhombic MGSIN2 from first-principles calculations. *Solid State Communications*, *336*, 114318.

[12]. Lee, C.-W.; Din, N. U.; Yazawa, K.; Brennecka, G. L.; Zakutayev, A.; Gorai, P. Emerging Materials and Design Principles for Wurtzite-Type Ferroelectrics. Matter 2024, 7 (4), 1644–1659.

[13]. Lee, C.-W.; Yazawa, K.; Zakutayev, A.; Brennecka, G. L.; Gorai, P. Switching It up: New Mechanisms Revealed in Wurtzite-Type Ferroelectrics. Science Advances 2024, 10 (20), eadl0848.

[14]. Kageyama, S.; Okamoto, K.; Yasuoka, S.; Ide, K.; Hanzawa, K.; Hiranaga, Y.; Hsieh, P.; Harza, S.; Suceava, A.; Saha, A.; Yokota, H.; Shigematsu, K.; Azuma, M.; Gopalan, V.; Uchida, H.; Hiramatsu, H.; Funakubo, H. Realization of Non-Equilibrium wurtzite structure in heterovalent ternary MGSIN2 film grown by reactive sputtering. Advanced Electronic Materials 2025. https://doi.org/10.1002/aelm.202400880.

[15]. Vangipuram, V. G. T.; Hu, C.; Majumder, A. M.; Chae, C.; Zhang, K.; Hwang, J.; Kash, K.; Zhao, H. Metal-organic chemical vapor deposition of MgSiN2 thin films. arXiv.org. https://arxiv.org/abs/2502.17306.

[16]. Wang, X., Che, S.-B., Ishitani, Y., & Yoshikawa, A. (2006). Effect of epitaxial temperature on n-Polar InN films grown by Molecular Beam Epitaxy. *Journal of Applied Physics*, *99*(7). https://doi.org/10.1063/1.2190720

[17]. Ooi, M. E., Ng, S. S., & Pakhuruddin, M. Z. (2025). Mocvd growth of inn thin films at different temperatures using pulsed trimethylindium approach. Journal of Alloys and Compounds, 1016, 178992.

[18]. Ratsch, C., & Venables, J. A. (2003). Nucleation theory and the early stages of thin film growth. Journal of Vacuum Science & Technology A: Vacuum, Surfaces, and Films, 21(5).

[19]. Yusof, A. S., Hassan, Z., Ould Saad Hamady, S., Ng, S. S., Ahmad, M. A., Lim, W. F., Che Seliman, M. A., Chevallier, C., & Fressengeas, N. (2021). The role of growth temperature on the indium incorporation process for the Mocvd growth of Ingan/Gan heterostructures. Microelectronics International, 38(3), 105–112.

[20]. Jumaah, O., & Jaluria, Y. (2019). The effect of carrier gas and reactor pressure on gallium nitride growth in MOCVD manufacturing process. Journal of Heat Transfer, 141(8).

[21]. Hu, S., Liu, S., Zhang, Z., Yan, H., Gan, Z., & Fang, H. (2015). A novel MOCVD reactor for growth of high-quality GaN-related led layers. Journal of Crystal Growth, 415, 72–77.

**Table Captions**

**Table 1:** Details of growth conditions for samples grown on GaN-on-sapphire templates.

**Table 2:** Sample A, K, D, E and F condition details and extracted optical bandgap. MgSiN$_2$ films is grown on c-plane, double-side polished sapphire substrates.

**Table 3:** Extracted EDX composition of elements for MgSiN$_2$ film from TEM films of Sample E. Mg:Si composition ratio of 0.92 obtained from EDX measurements.

**Figure Captions**

**Figure 1:** The measured Mg/Si atomic ratios for MgSiN$_2$ films grown on GaN templates as a function of the Mg/Si precursor molar flow rate ratio (FRR) for different growth temperatures and chamber pressures.

**Figure 2:** XRD ω–2θ scans (left), AFM topographies (center), and SEM micrographs (right) for samples A, B, and C grown on GaN-on-sapphire at 900 °C with varying Mg:Si molar FRRs and NH$_3$ flow rates.

**Figure 3:** XRD ω–2θ scans (left), AFM images (center), and SEM micrographs (right) for samples A, D, E, and F, each grown on GaN-on-sapphire under different Mg:Si molar FRR and NH$_3$ flow conditions.

**Figure 4:** XRD ω–2θ scans (left), AFM images (center), and SEM micrographs (right) for samples G, E, H, I, and J grown under different Mg:Si molar FRR at 945°C and 500 Torr.

**Figure 5:** XRD ω–2θ scans (left), AFM images (center), and SEM micrographs (right) for samples K, L, M, N, and O grown under different Mg:Si molar FRR at 945°C and 450 Torr.

**Figure 6:** XRD ω–2θ scans (left), AFM topography (center), and SEM micrographs (right) for samples P, Q, and R grown under different Mg:Si molar FRR at 945°C and 550 Torr.

**Figure 7:** Rocking curve of the (002) MgSiN$_2$ peak for sample I, highlighting its full width at half maximum (FWHM). A relatively narrow FWHM suggests good crystalline alignment and low defect density, consistent with the smooth surface morphology and sharp diffraction peaks observed in other characterizations of this sample.

**Figure 8:** (a) Transmittance spectra of several MgSiN$_2$ films grown on double-side polished sapphire, illustrating a pronounced absorption edge in the deep-UV range (5.5–6.5 eV). (b) Corresponding Tauc plots, where the linear extrapolation of $(\alpha h\nu)^2$ versus $h\nu$ is used to estimate the direct band gap. (c) Corresponding Tauc plots, where the linear extrapolation of $(\alpha h\nu)^{1/2}$ versus $h\nu$ is used to estimate the indirect band gap.

**Figure 9:** (a) and (b) STEM cross-section of sample E showing the MgSiN$_2$ film grown on a GaN-on-sapphire template at two different length scales; (c) the STEM cross-section across a larger area for EDX mapping; (d) elemental EDX mapping for (i) Ga, (ii) Si, (iii) N and, (iv) Mg; (e) quantitative EDX elemental mapping across the MgSiN$_2$ film from the MgSiN$_2$ surface to the interface with the GaN template.

**Figure 10:** (a) Simulated diffraction pattern for the wurtzite GaN structure; (b) diffraction pattern of the GaN obtained from the STEM cross-section of sample E; (c) simulated diffraction pattern for the orthorhombic MgSiN$_2$ structure; (d) diffraction pattern of the MgSiN$_2$ film obtained from the STEM cross-section of sample E.

**Table 1**

| Sample ID | Growth ID | Growth Temperature (°C) | Growth Pressure (Torr) | Mg:Si Precursor Molar FRR | $NH_3$ flow (slm) | Measured SEM-EDX Mg:Si Ratio | RMS (nm) |
|---|---|---|---|---|---|---|---|
| A | MgSiN#061 | 900 | 500 | 9 | 2.5 | 1.12 ± 0.10 | 24.2 ± 1.2 |
| B | MgSiN#062 | 900 | 500 | 9 | 3.5 | 1.27 ± 0.12 | 40.2 ± 2.0 |
| C | MgSiN#063 | 900 | 500 | 7.5 | 3.5 | 1.09 ± 0.10 | 26.2 ± 1.3 |
| D | MgSiN#069 | 928 | 500 | 11.3 | 2.5 | 1.03 ± 0.10 | 16.2 ± 0.8 |
| E | MgSiN#070 | 945 | 500 | 12.3 | 2.5 | 0.99 ± 0.09 | 5.0 ± 0.3 |
| F | MgSiN#071 | 960 | 500 | 13.5 | 2.5 | 0.96 ± 0.09 | 6.5 ± 0.4 |
| G | MgSiN#075 | 945 | 500 | 10.5 | 2.5 | 0.57 ± 0.05 | 4.4 ± 0.2 |
| H | MgSiN#078 | 945 | 500 | 14.2 | 2.5 | 0.98 ± 0.08 | 9.0 ± 0.5 |
| I | MgSiN#084 | 945 | 500 | 15.1 | 2.5 | 1.02 ± 0.10 | 1.5 ± 0.1 |
| J | MgSiN#087 | 945 | 500 | 15.6 | 2.5 | 1.00 ± 0.10 | 5.3 ± 0.3 |
| K | MgSiN#072 | 945 | 450 | 12.3 | 2.5 | 0.78 ± 0.08 | 6.2 ± 0.3 |
| L | MgSiN#074 | 945 | 450 | 13.5 | 2.5 | 0.86 ± 0.08 | 4.4 ± 0.2 |
| M | MgSiN#082 | 945 | 450 | 14.5 | 2.5 | 1.02 ± 0.10 | 4.6 ± 0.2 |
| N | MgSiN#077 | 945 | 450 | 15.6 | 2.5 | 0.96 ± 0.08 | 3.1 ± 0.2 |
| O | MgSiN#083 | 945 | 450 | 16.6 | 2.5 | 1.02 ± 0.10 | 12.1 ± 0.6 |
| P | MgSiN#091 | 945 | 550 | 12.9 | 2.5 | 1.08 ± 0.10 | 9.6 ± 0.5 |
| Q | MgSiN#090 | 945 | 550 | 13.8 | 2.5 | 1.05 ± 0.10 | 6.5 ± 0.3 |
| R | MgSiN#088 | 945 | 550 | 15.1 | 2.5 | 1.02 ± 0.10 | 5.9 ± 0.3 |

**Table 2**

| Sample ID | Growth ID | Measured SEM-EDX Mg:Si Ratio | RMS (nm) | Extracted direct Bandgap, $E_{direct}$ (eV) | Extracted Indirect Bandgap, $E_{indirect}$ (eV) |
|---|---|---|---|---|---|
| A | MgSiN#061 | 1.12 ± 0.10 | 24.2 ± 1.2 | 6.27 ± 0.03 | 5.40 ± 0.03 |
| K | MgSiN#072 | 0.78 ± 0.08 | 6.2 ± 0.3 | 6.28 ± 0.03 | 5.54 ± 0.03 |
| D | MgSiN#069 | 1.03 ± 0.10 | 16.2 ± 0.8 | 6.35 ± 0.03 | 5.77 ± 0.03 |
| E | MgSiN#070 | 0.99 ± 0.09 | 5.0 ± 0.3 | 6.38 ± 0.03 | 5.81 ± 0.03 |
| F | MgSiN#071 | 0.96 ± 0.09 | 6.5 ± 0.4 | 6.36 ± 0.03 | 5.78 ± 0.03 |

**Table 3**

| Extracted EDX Composition of Sample E from TEM | |
|---|---|
| **Element** | **Atomic (%)** |
| Mg | 25 ± 4 |
| Si | 27 ± 4 |
| N | 47 ± 3 |
| Ga | 0.54 ± 0.08 |

**Figure 1**

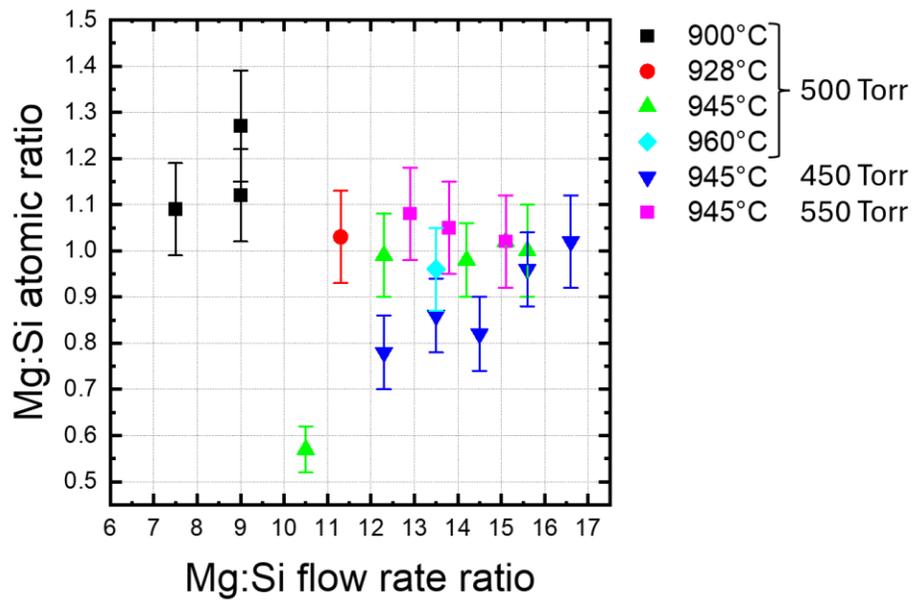

**Figure 2**

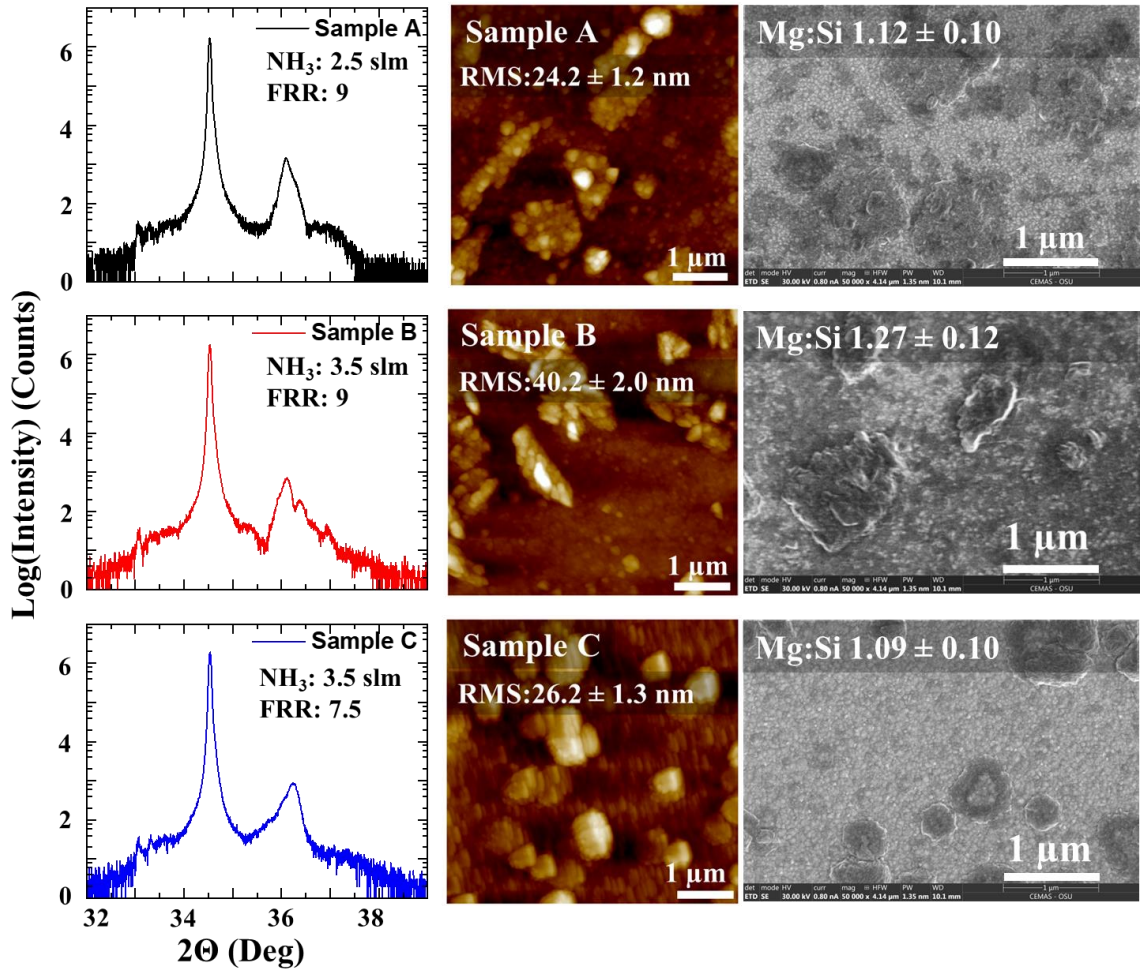

**Figure 3**

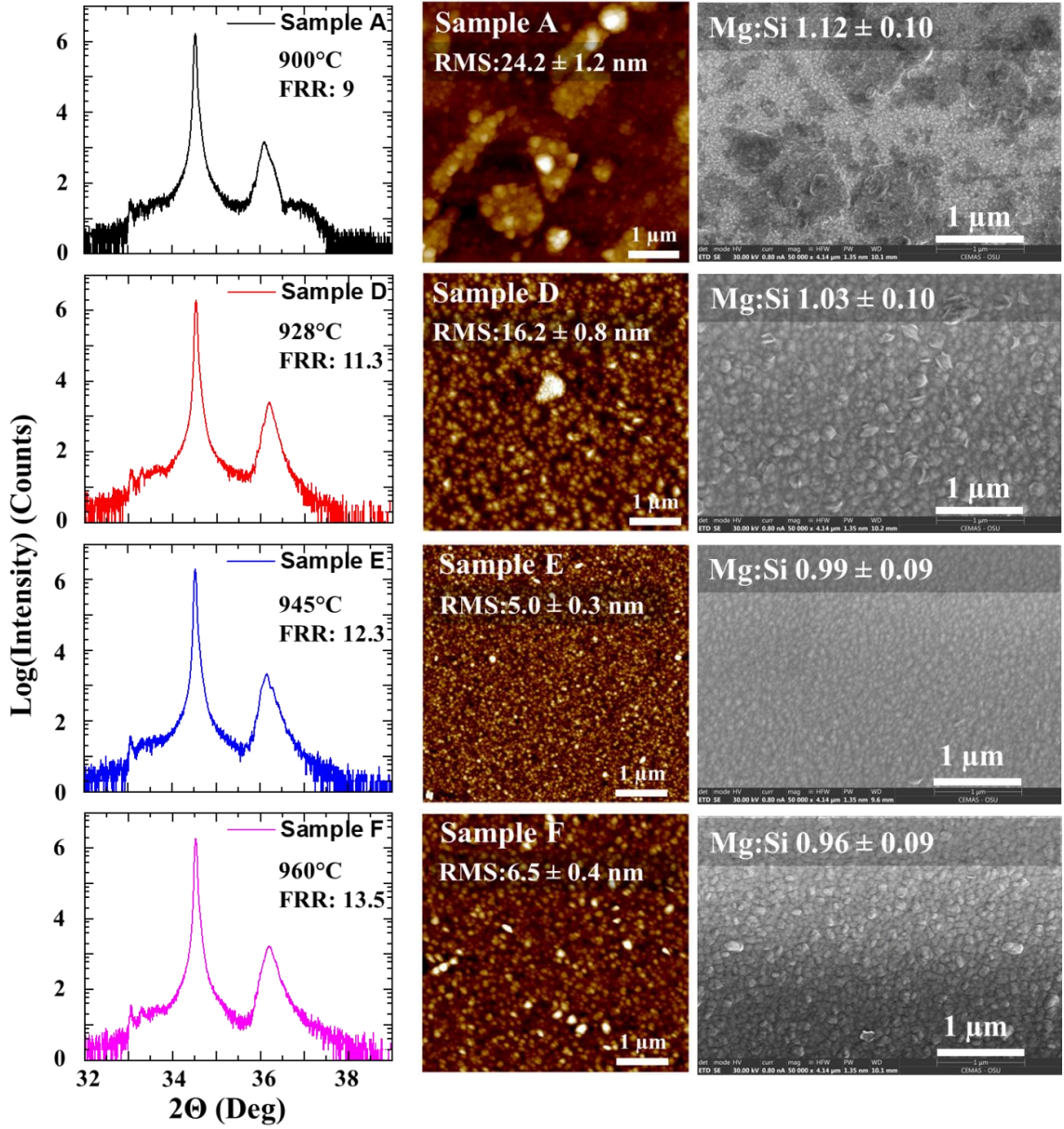

**Figure 4**

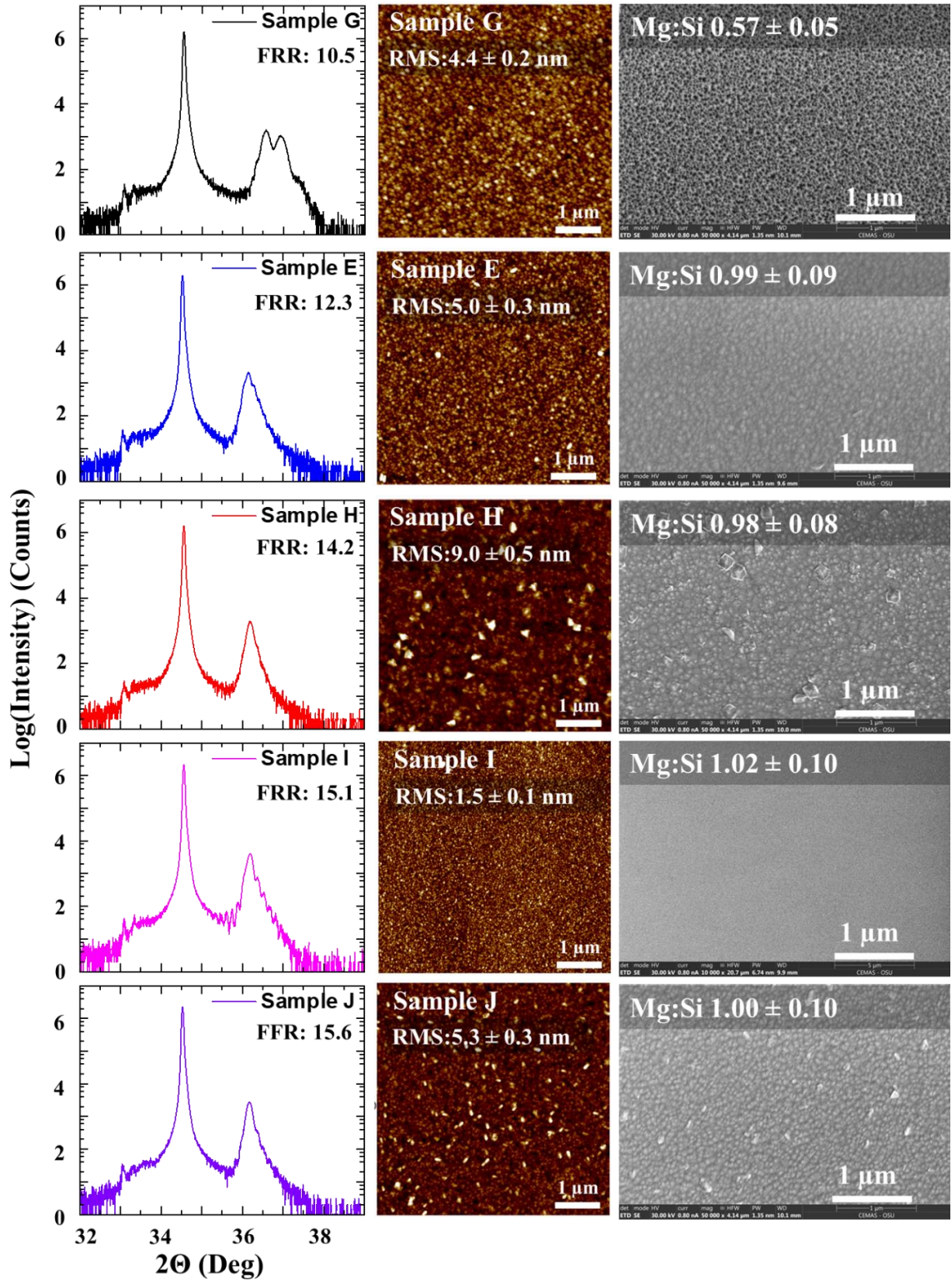

**Figure 5**

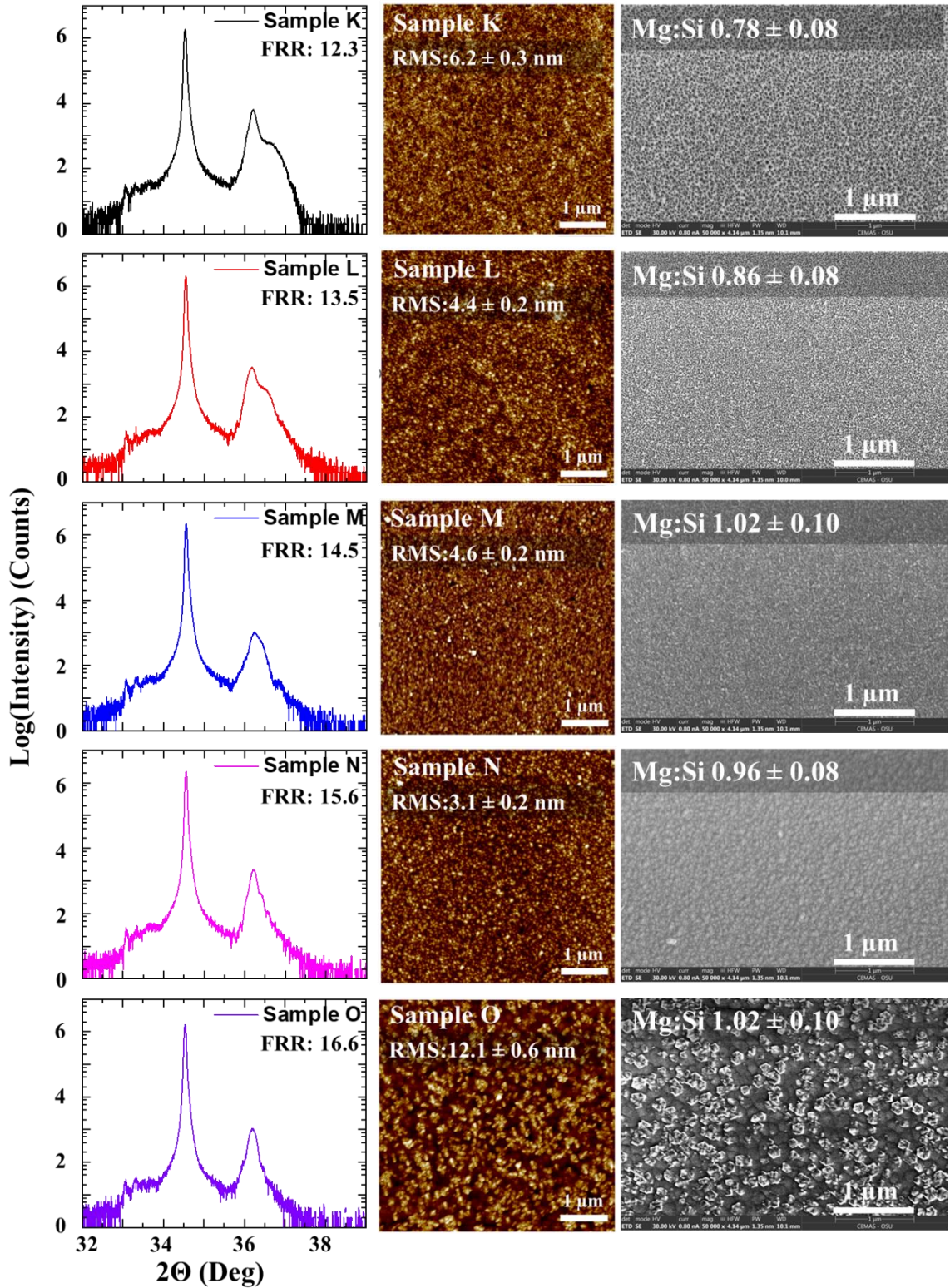

**Figure 6**

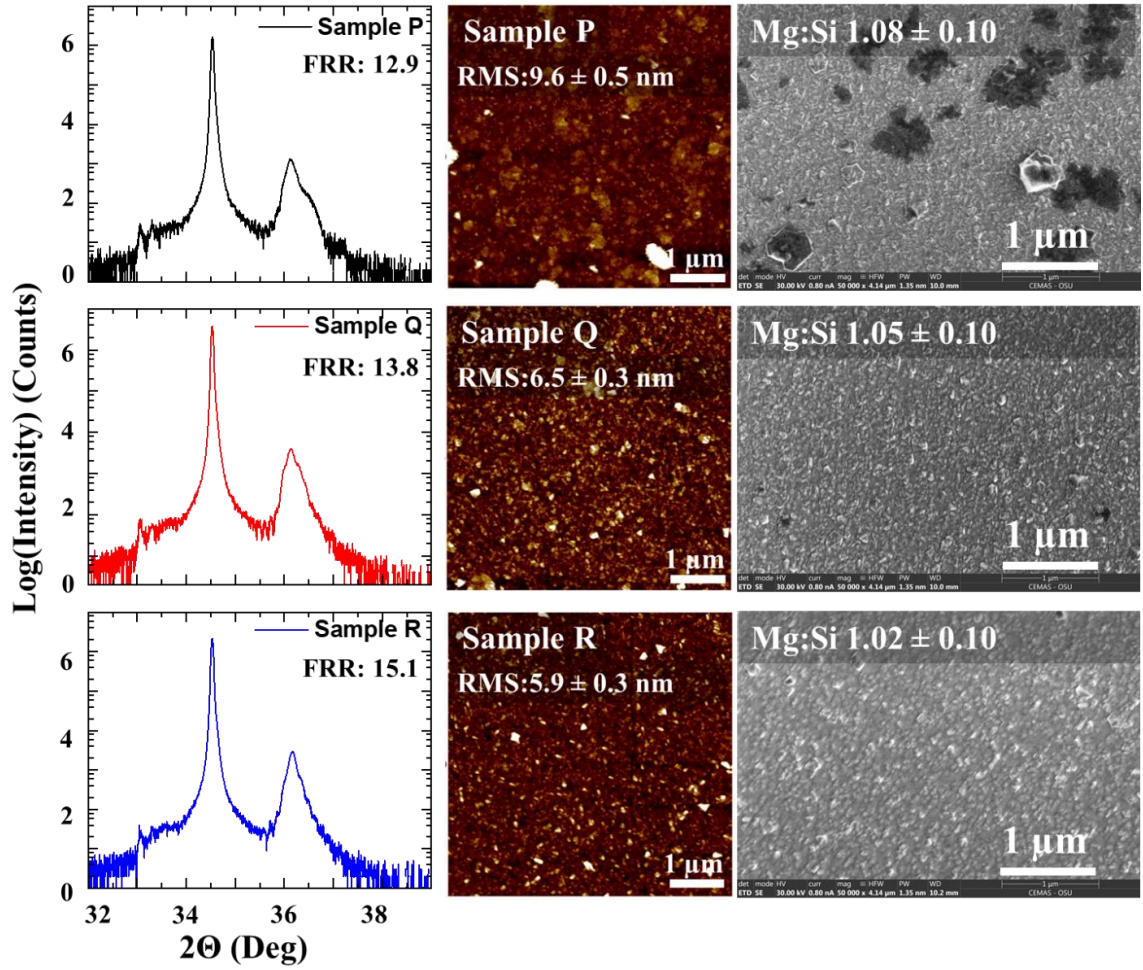

**Figure 7**

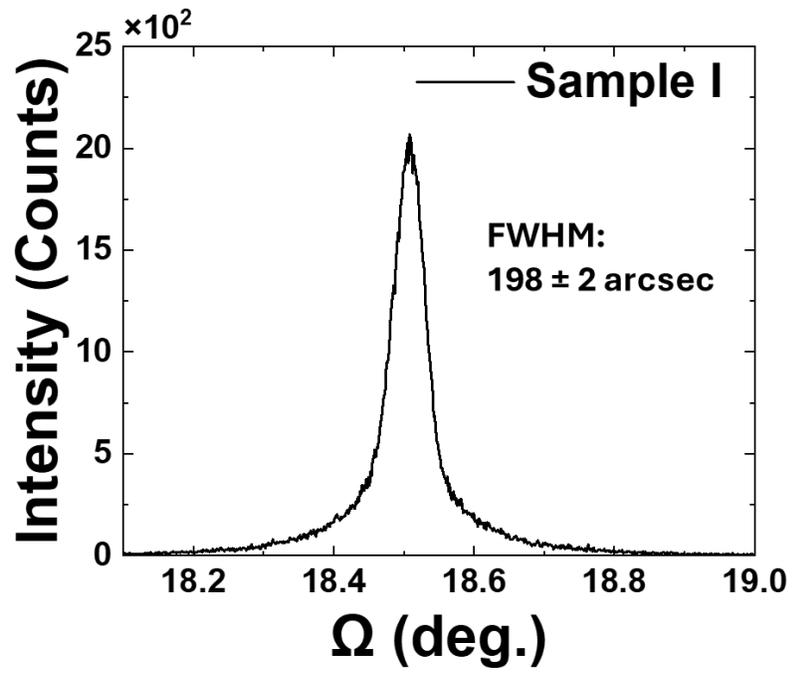

**Figure 8**

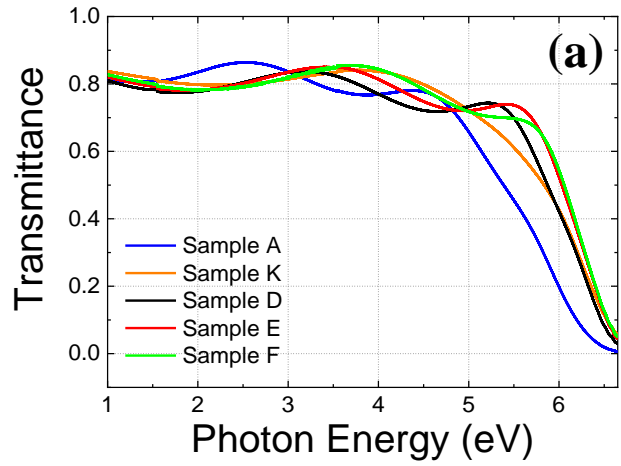

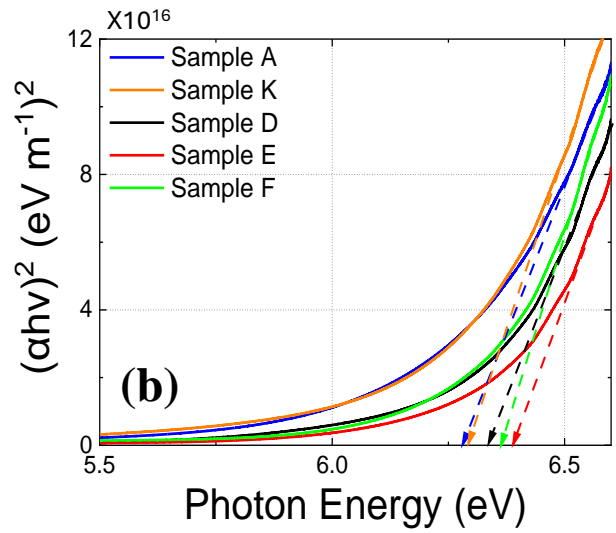

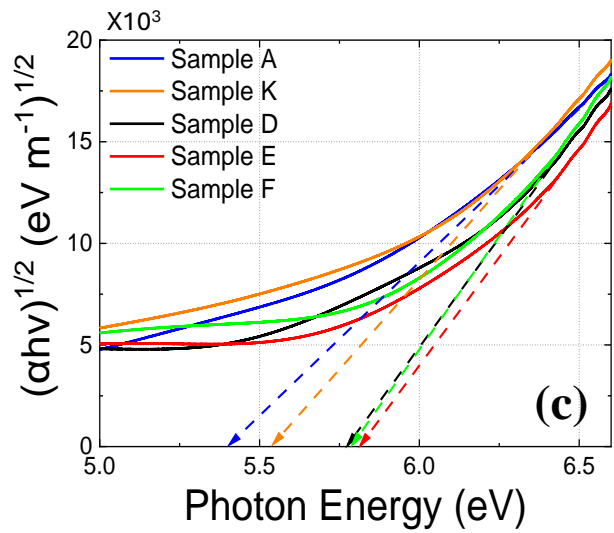

**Figure 9**

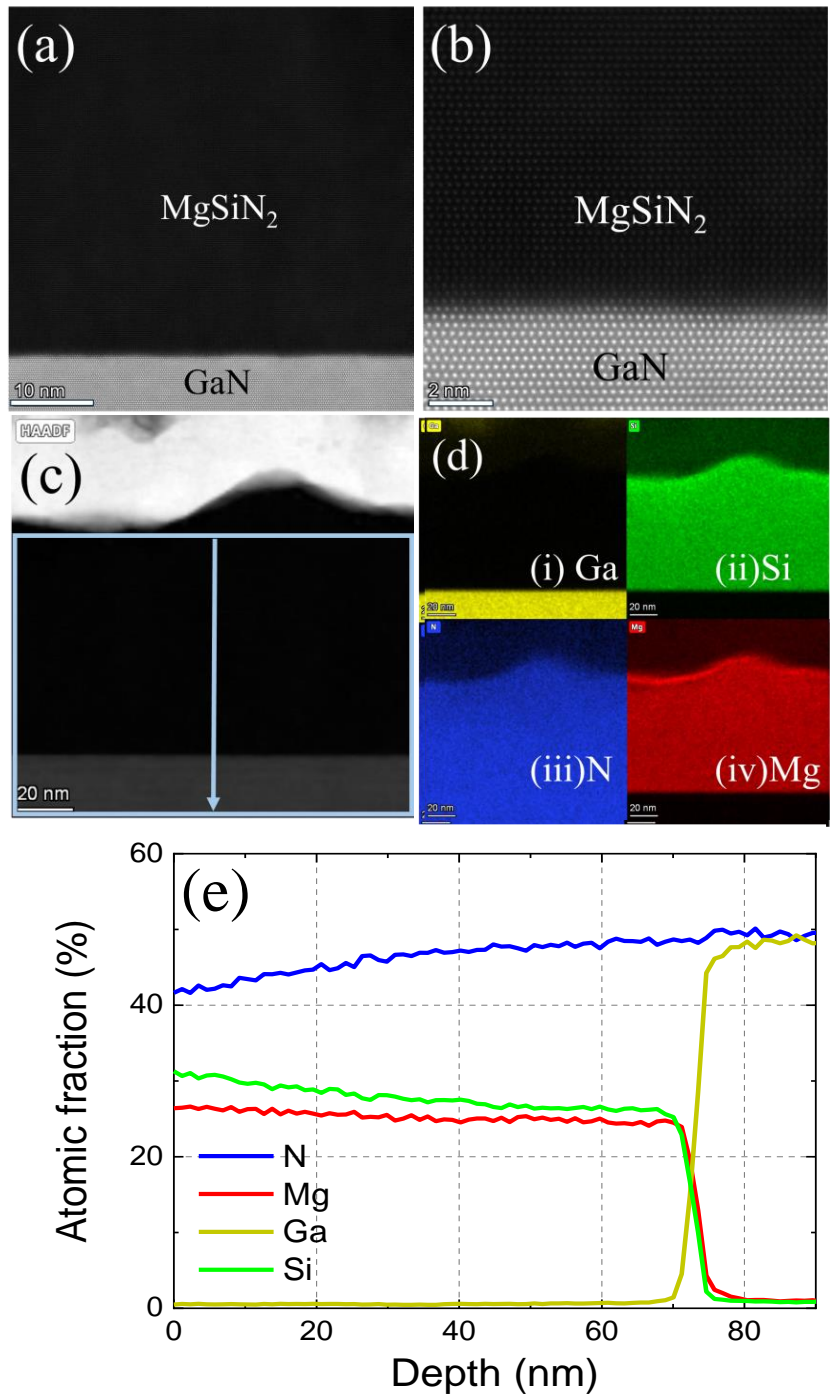

**Figure 10**

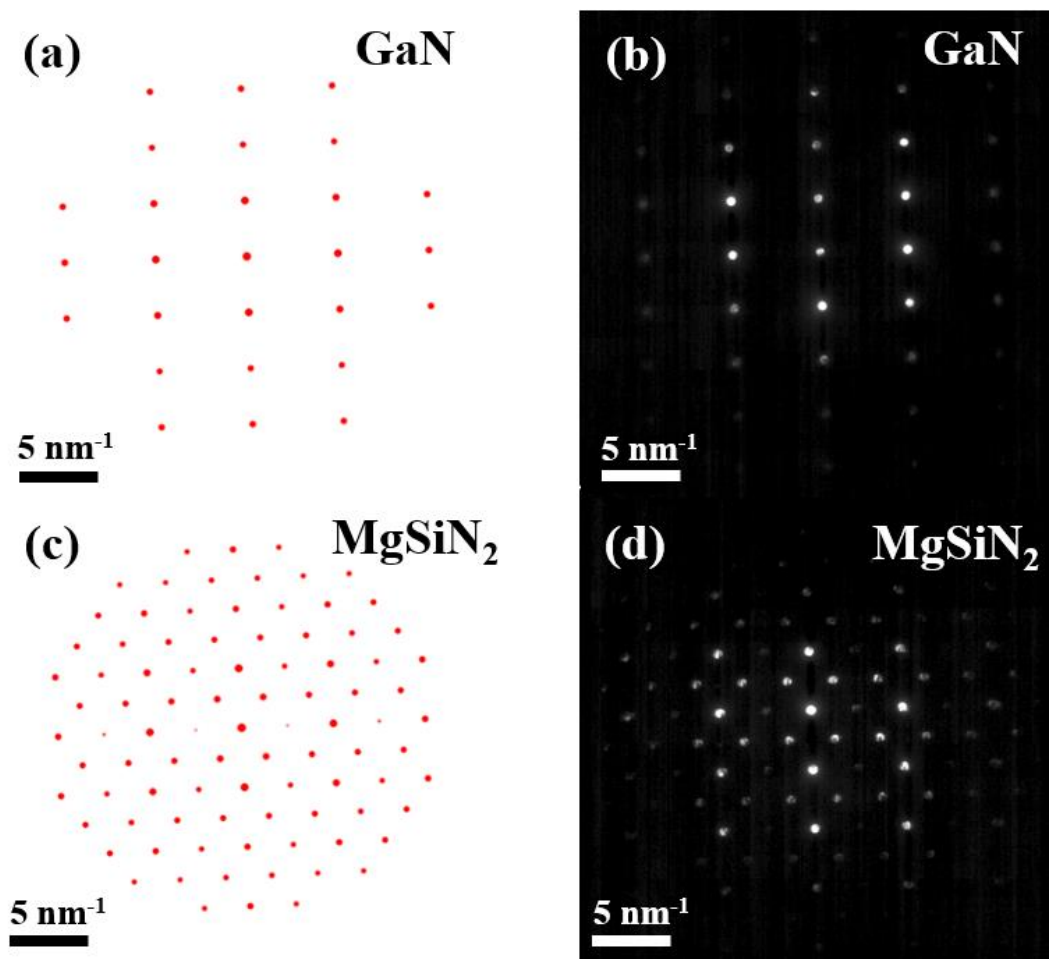